\documentclass[journal]{IEEEtran}

\usepackage{epsfig,amsmath,amssymb,epsf,amsthm,scalefnt,multirow,subfig}
\usepackage{graphics,amsmath,amssymb,graphicx,amsthm,scalefnt,multirow,dsfont,subfig}
\usepackage{algorithm}
\usepackage{algpseudocode}
\usepackage{xcolor}
\usepackage{float}
\usepackage{cite}
\usepackage{psfrag}
\usepackage{gensymb}
\usepackage{cleveref}
\usepackage{enumitem}
\usepackage{amsthm}

\newtheorem*{lemma*}{Lemma}

 \def\cF{{\mathcal{F}}}  
   
   \def\cP{{\mathcal{P}}}
\def\cQ{{\mathcal{Q}}}

\def\diag{\mathop{\mathrm{diag}}}

\def\b0{{\pmb{0}}} 

\def\ba{{\mathbf{a}}}   
   \def\bh{{\mathbf{h}}}
   
   \def\bp{{\mathbf{p}}}
  \def\bs{{\mathbf{s}}} 
  \def\bw{{\mathbf{w}}} 
\def\by{{\mathbf{y}}}   

\def\bA{{\mathbf{A}}}   
   \def\bH{{\mathbf{H}}}
\def\bI{{\mathbf{I}}}   
   
  \def\bS{{\mathbf{S}}} 
  \def\bW{{\mathbf{W}}} 
\def\bY{{\mathbf{Y}}}

\begin{document}

\title{{\huge Downlink Extrapolation for FDD Multiple Antenna Systems Through Neural Network Using Extracted Uplink Path Gains}}

\author{Hyuckjin Choi,~\IEEEmembership{Student Member,~IEEE} and Junil Choi,~\IEEEmembership{Member,~IEEE}
	\thanks{H. Choi and J. Choi are with the School of Electrical Engineering, Korea Advanced Institute of Science and Technology (e-mail: \{hugzin008, junil\}@kaist.ac.kr).}}  
\maketitle

\begin{abstract}
When base stations (BSs) are deployed with multiple antennas, they need to have downlink (DL) channel state information (CSI) to optimize downlink transmissions by beamforming. The DL CSI is usually measured at mobile stations (MSs) through DL training and fed back to the BS in frequency division duplexing (FDD). The DL training and uplink (UL) feedback might become infeasible due to insufficient coherence time interval when the channel rapidly changes due to high speed of MSs. Without the feedback from MSs, it may be possible for the BS to directly obtain the DL CSI using the inherent relation of UL and DL channels even in FDD, which is called DL extrapolation. Although the exact relation would be highly nonlinear, previous studies have shown that a neural network (NN) can be used to estimate the DL CSI from the UL CSI at the BS. Most of previous works on this line of research trained the NN using full dimensional UL and DL channels; however, the NN training complexity becomes severe as the number of antennas at the BS increases. To reduce the training complexity and improve DL CSI estimation quality, this paper proposes a novel DL extrapolation technique using simplified input and output of the NN. It is shown through many measurement campaigns that the UL and DL channels still share common components like path delays and angles in FDD. The proposed technique first extracts these common coefficients from the UL and DL channels and trains the NN only using the path gains, which depend on frequency bands, with reduced dimension compared to the full UL and DL channels. Extensive simulation results show that the proposed technique outperforms the conventional approach, which relies on the full UL and DL channels to train the NN, regardless of the speed of MSs.
\end{abstract}


\section{Introduction}
\label{sec:introduction}
Time division duplexing (TDD) is becoming popular due to its flexibility of supporting imbalanced uplink (UL) and downlink (DL) traffics and low overhead on channel state information (CSI) acquisition at base stations (BSs) when the number of antennas at the BS is large \cite{Larsson:2014}. Still, frequency division duplexing (FDD) is important in fifth generation (5G) and beyond cellular systems to support backward compatibility. The conventional way for the BS to obtain the DL CSI in FDD is based on DL training and UL limited feedback \cite{Love:2008}. The training and feedback overhead, however, increases with the number of antennas at the BS and would become excessive in massive multiple-input multiple-output (MIMO) systems \cite{hassibi2003much,jindal2006mimo,Larsson:2014,Marzetta:2015}.

There have been many works to resolve the DL training and UL feedback issues in massive MIMO. For both issues, most of previous works relied on long term channel statistics to reduce the training and feedback overhead \cite{JChoi:2013,Adhikary:2013,JChoi:2014,Noh:2014,Gao:2015,Sun:2015,Gao:2016}.
Although effective, these approaches are vulnerable to the lack of coherence time interval, which usually happens when mobile stations (MSs) travel with high speed. Recently, there has been a line of research to completely or partially get rid of the DL training and UL feedback overhead in FDD by DL extrapolation using the UL CSI \cite{Fran:2019,Arn:2019,Alrabeiah:2019,Yang:2019,vasisht2016eliminating}.

The DL extrapolation exploits inherent UL and DL channel reciprocity that occurs even in FDD, e.g., dominant UL and DL path angles, since the UL and DL signals traverse the same environment. Although some UL and DL channel parameters experience the reciprocity in FDD, it is critical to define physical relation between the two channels in different frequencies for the DL extrapolation to work. It is very difficult, however, to show the relationship mathematically because the relationship is governed by the Maxwell's equations with practical environments, e.g., buildings, trees, or vehicles, as boundary conditions, which makes the relationship highly nonlinear.

Deep learning using a neural network (NN) can figure out complex relation between two data sets. Recently, deep learning has been applied to many wireless communication problems including beamforming, limited feedback, and modulation \cite{Xiao:2019,Wen:2018,Elbir:2019,Huang:2019,Wang:2019}. When using the NN for the DL extrapolation, most of previous works exploited the full dimensional UL and DL channels as the input and output of the NN using simplified analytical relation of UL and DL channels \cite{Arn:2019,Alrabeiah:2019, Yang:2019}. It may be not possible, however, to analytically define the relation between the UL and DL channels in practice. Moreover, these works did not consider any MS mobility. Using the full UL and DL channels for the input and output of the NN might confuse the NN to find their complex relation when the MS moves with high speed.

To reduce the NN learning complexity, we propose to first extract angles of departure (AoDs), delays, and path gains, and exploit the path gains obtained from similar signal path angles in the UL and DL to train the NN. Although the path gains are still governed by the Maxwell's equations, the information each path gain contains is much more compact than the channel itself, making the NN work even better.

After the training, the NN gives predicted DL path gains directly from the extracted UL path gains. With the partial information of channels already obtained from the UL CSI, i.e., the delays and AoDs that experience the reciprocity in FDD, the BS can reconstruct the DL channel using the DL path gains. After the UL and DL path gain extraction, the dimension of the input and output of the NN can become much smaller compared to the full UL and DL channels. The reduced input and output dimension enables to have smaller number of weights connecting the NN nodes as well as to simplify the weight updating process, which makes it possible to achieve the same or improved accuracy with reduced NN training time compared to using the full UL and DL channels.

The DL extrapolation techniques would be most beneficial when the MSs move with high speed. The conventional DL training-based approach in FDD, which requires the large training overhead, becomes infeasible when channels vary fast due to high mobility. The DL extrapolation techniques using the trained NNs have much smaller processing time for the BS to acquire the DL CSI than the DL training; therefore, the training overhead can be dramatically reduced to ensure sufficient time for data transmissions. The proposed DL extrapolation technique makes the NN simpler, even make it more suitable to the environments suffering from a short coherence time. The main contributions are summarized in below:

\begin{enumerate}
	\item We train the network with the realistic channel samples generated by QuaDRiGa, a realistic channel emulator \cite{QuaDRiGa:2014}, which means the channel samples used for the numerical validation are not obtained from a pre-defined channel model. By using QuaDRiGa, we can evaluate our scheme with realistic data, which consider various physics, e.g., carrier frequency, mobility, and total bandwidth, and ensure the practicality of proposed method.
	\item We propose the path gain extraction method to simplify the input and output of NN for the DL extrapolation. Just using the path gains for the learning instead of the full dimensional channels, the training can be more rapidly and precisely conducted.
	\item By properly taking DL extrapolation processing time into account for the NN training, the proposed NN-based DL extrapolation performs well even when MSs move with high speed. 
\end{enumerate}

The organization of paper is as follows. In Section \ref{System Model}, we introduce the process of generating channel samples using QuaDRiGa and the channel model we use for the proposed DL extrapolation technique. Conventional ways of obtaining the DL CSI through the DL training and NN-based DL extrapolation are presented in Section \ref{DL Acq}. In Section \ref{Param extrac}, we explain channel parameter extraction algorithms for the proposed NN-based DL extrapolation, and detailed NN settings are described in Section \ref{NN Settings}. Extensive numerical results are given in Section \ref{Numerical result}, and the conclusion follows in Section \ref{conclusion}.

\textbf{Notation:} Lower and upper boldface letters denote column vectors and matrices. $\bA^\ast$, $\bA^\mathrm{T}$, $\bA^{-1}$ and $\bA^\mathrm{H}$ are used to represent the conjugate, transpose, inverse, and Hermitian (conjugate transpose) of the matrix $\bA$. $\Vert\cdot\rVert_2$ and $\lVert\cdot\rVert_\mathrm{F}$ denotes the $\ell_2$-norm of the complex vector and Frobenius norm of the complex matrix. $\bI_N$ denotes the $N\times N$ identity matrix. $\{\alpha_n\}^N$ represents a set of $\alpha_n$ for $n=1,\dots,N$, and $\{\alpha_{m,n}\}^{M,N}$ denotes a set of $\alpha_{m,n}$ for $(m,n)=(1,1),\dots,(M,N)$. $\diag[a_1,\dots,a_N]$ is the $N\times N$ diagonal matrix where its diagonal entries are $a_1,\dots,a_N$.

\begin{figure}
	\centering
	\includegraphics[width=1.05\columnwidth]{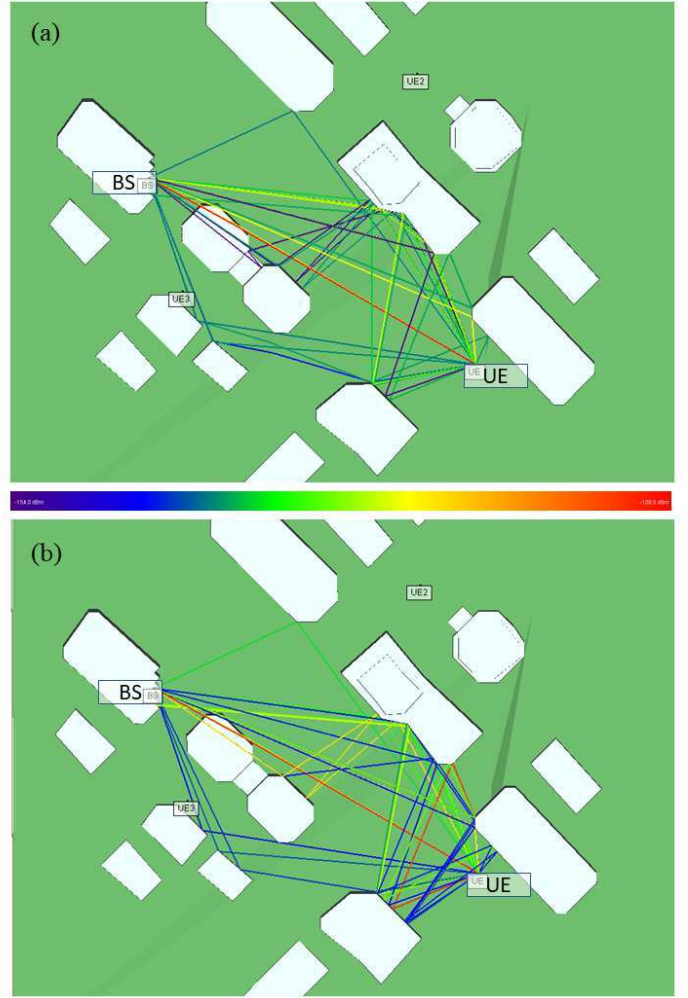}
	\caption{Channels generated by Wireless InSite \cite{WirelessInSite} in an urban area. (a) Uplink channel with 2.14 GHz carrier frequency and (b) downlink channel with 1.95 GHz carrier frequency.}
	\label{InSiteChannel}
\end{figure}

\section{System Model}\label{System Model}

We first describe the details of generating channel samples using QuaDRiGa for the NN in Section \ref{Gen_ch}. Then, to develop the proposed DL extrapolation technique, we define the channel model, which approximates the channel generated by QuaDRiGa, in Section \ref{Channel Model}.

\subsection{Generating Channel Samples}\label{Gen_ch}

We consider a scenario that the NN at the BS is trained with the UL and DL CSI from a given path where MSs travel and conduct the DL extrapolation for a new MS entering the same path. The DL extrapolation is possible as long as the UL and DL experience the channel reciprocity in FDD. Figs. \ref{InSiteChannel}-(a) and (b) are the snapshots of channels in different carrier frequencies, which are generated by Wireless InSite, a commercial ray-tracing simulator \cite{WirelessInSite}. As shown in the figures, the traces of UL signal, with 2.14 GHz carrier frequency, and DL signal, with 1.95 GHz carrier frequency, are almost identical even with the guard band between the two.

\begin{figure}
	\centering
	\includegraphics[width=1\columnwidth]{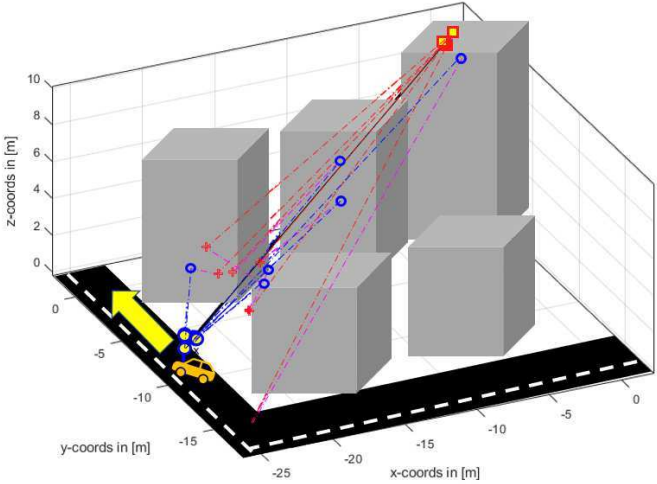}
	\caption{A typical urban communication environment. The track of MS is restricted on a given path. Red lines visualize the trajectory of signals from the BS to the first bounce scatter. Blue lines visualize the trajectory of signals from the last bounce scatter to the MS.}
	\label{givenPath}
\end{figure}

To exploit the NN for DL extrapolation, the BS should collect the CSI data from MSs moving on the same path. A typical scenario of an urban communication system is depicted in Fig. \ref{givenPath}. Since urban roads restrict the course of MSs as shown in the figure, the tracks of MSs on a given path can be assured to be nearly identical. We exploit QuaDRiGa \cite{QuaDRiGa:2014} to obtain samples of the UL and DL channels since it considers complicated physics including mobility for channel emulation. The channel samples obtained from QuaDRiGa does not have any idealistic feature, e.g., perfect directional reciprocity between the UL and DL channels. This is quite different from previous works where channel samples are obtained based on analytical channel models \cite{Alrabeiah:2019,Yang:2019,Xiao:2019,Dong:2019,Wen:2018,Elbir:2019,Huang:2019,Wang:2019,Viera:2017}.

QuaDRiGa periodically generates channel snapshots for each MS moving on a given path. The MS sample set is a set of channel snapshots of one MS along the given path. The environment geometry on a given path, e.g., nearby buildings, is maintained for a long time. Therefore, the MSs on the same path would experience high spatial correlation. However, specific signal paths may vary temporarily due to large dynamic objects like moving trucks. Taking this situation into an account, for each MS sample set, we fix several clusters to reflect the static environments and randomly change a few clusters to reflect the dynamic objects. The UL and DL channels for each MS have the same clusters to maintain their geometrical reciprocity. 

\textit{\textbf{Remark 1:}} Even with the trained NN, it takes a time for the BS to conduct the DL extrapolation. When the BS estimates the UL CSI at $t_0$, it needs to infer the DL CSI at $t_0+d$ where $d$ is the processing delay for the DL extrapolation. Thus, we generate UL and DL channels to keep their time interval $d$.

\textit{\textbf{Remark 2:}} To implement the DL extrapolation using the NN, the BS has to collect the UL and DL CSI as training data sets. In the conventional limited feedback for FDD, the BS can only access to quantized DL CSI. Still, we believe it is possible to train the NN with high resolution DL CSI once services providers obtain the required data sets offline since the NN training is not a real-time process. Although the exact timestamps of UL and DL CSI data may be different, the timing difference can be adjusted before the training process since the training is performed after stacking the data. After the services providers train the NN using collected UL and DL CSI, the BS only needs to operate the DL extrapolation function in real-time using the trained NN stored in its memory.

\subsection{Channel Model}\label{Channel Model}

The channel generated by QuaDRiGa can be approximated as
\begin{align}\label{channel model1}
\bh(t,f_\mathrm{c})=\sum_{\ell=1}^L\bh_\ell(t,f_c)=\sum_{\ell=1}^L\sum_{p=1}^P\alpha_{\ell,p}(f_\mathrm{c})\delta(t-\tau_\ell)\ba(\theta_{\ell,p}),
\end{align}
which consists of $L$ paths and $P$ subpaths per path where $L$ represents the number of clusters in our case. In \eqref{channel model1}, $f_\mathrm{c}$ is the carrier frequency, $\alpha_{\ell,p}(f_\mathrm{c})$ is the path gain, $\tau_\ell$ is the delay, and $\theta_{\ell,p}$ is the AoD. We assume the path gains and AoDs are subpath dependent while all subpaths from the same cluster experience the same delay. The $N_\mathrm{BS}\times 1$ vector $\ba(\theta_{\ell,p})$ is the array response of the signal incident at $\theta_{\ell,p}$, where $N_\mathrm{BS}$ is the number of BS antennas.

The channel due to the $\ell$-th cluster is given as
\begin{align}\label{h_l}
\bh_\ell(t,f_c)=\sum_{p=1}^P\alpha_{\ell,p}(f_c)\delta(t-\tau_\ell)\ba(\theta_{\ell,p})=\boldsymbol{\mathfrak{h}}_\ell(f_c)\delta(t-\tau_\ell).
\end{align}
The channel coefficient of the $\ell$-th cluster $\boldsymbol{\mathfrak{h}}_\ell(f_c)$ consists of $P$ subpath components; $\boldsymbol{\mathfrak{h}}_\ell(f_c)=\sum_{p=1}^P\alpha_{\ell,p}(f_c)\ba(\theta_{\ell,p})$.
While the path gains are frequency dependent components, we assume the delays and AoDs are frequency independent components as in \cite{Fran:2019}, i.e., the UL and DL channels experience the same delays and AoDs due to the same geometry.

The channel in \eqref{channel model1} can be transformed into the frequency domain. The $k$-th subcarrier of the orthogonal frequency division multiplexing (OFDM) channel is obtained as \cite{Fran:2019}
\begin{align}\label{channel model2}
\underline{\bh}_k(f_\mathrm{c}) &= \cF_k\{\bh(t,f_c)\} = \int_{-\infty}^\infty\bh(t,f_c)e^{-j2\pi f_s\frac{k}{K}t}dt\notag \\
&=  \sum_{\ell=1}^L\sum_{p=1}^P\alpha_{\ell,p}(f_\mathrm{c})e^{-j2\pi f_s\tau_\ell\frac{k}{K}}\ba(\theta_{\ell,p})\notag \\
&= \sum_{\ell=1}^L\boldsymbol{\mathfrak{h}}_\ell(f_c)e^{-j2\pi f_s\tau_\ell\frac{k}{K}},
\end{align}
where $f_\mathrm{s}$ is the total bandwidth, and $K$ is the total number of subcarriers. We use the underline to denote variables in the frequency domain throughout the paper. By concatenating all subcarriers, we have the $N_\mathrm{BS}\times K$ OFDM channel matrix $\underline{\bH}(f_c)=[\underline{\bh}_1(f_c)\cdots\underline{\bh}_K(f_c)]$. The OFDM channel matrix $\underline{\bH}(f_c)$ can be rewritten as
\begin{align}\label{OFDMchannel}
\underline{\bH}(f_c)=\sum_{\ell=1}^L\boldsymbol{\mathfrak{h}}_\ell(f_c)\bp^\mathrm{T}(\tau_\ell),
\end{align}
where $\bp(\tau_\ell)=[e^{-j2\pi f_s\tau_\ell\frac{1}{K}},e^{-j2\pi f_s\tau_\ell\frac{2}{K}},\dots,e^{-j2\pi f_s \tau_\ell}]^\mathrm{T}$. Each element of $\bp(\tau_\ell)$ matches with the subcarrier of the OFDM channel.

\section{Review of DL CSI Acquisition}\label{DL Acq}

The conventional way of obtaining the DL CSI in FDD relies on the DL training and UL feedback, which cause large overhead especially for massive MIMO \cite{hassibi2003much}. In Section \ref{DLtrain}, we first explain the conventional DL training, which works as a baseline of the DL extrapolation techniques. Then, we briefly explain the previous DL extrapolation using the NN based on the full UL and DL channels in Section \ref{DLnn} while we detail the proposed technique in Section \ref{Param extrac}.

\subsection{Conventional DL Training}\label{DLtrain}
In the DL training, the MS estimates the DL channel and feeds back the acquired information to the BS. First, the BS transmits the pilot symbol $\bs_\mathrm{dl}(t)$ to the MS. The received signal at the MS in the time domain is written as 
\begin{align}\label{rxsig}
y_j(t,f_\mathrm{dl})=\sum_{\ell=1}^L\sum_{p=1}^P\alpha_{\ell,p}(f_\mathrm{dl})\bs_{\mathrm{dl},j}^\mathrm{H}(t-\tau_\ell)\ba(\theta_{\ell,p})+w_{\mathrm{dl},j}(t).
\end{align}
where $\bs_{\mathrm{dl},j}(t)$ is the $j$-th pilot symbol with $j\in\{1,\dots,J\}$, generally, $J\geq N_\mathrm{BS}$. The additive white Gaussian noise $w_{\mathrm{dl},j}(t)$ has zero mean with variance $N_0$. To estimate the DL CSI, the MS first stacks the $J$ received pilot signals as 
\begin{align}
\by(t,f_\mathrm{dl})=\sum_{\ell=1}^L\sum_{p=1}^P\alpha_{\ell,p}(f_\mathrm{dl})\bS_\mathrm{dl}^\mathrm{H}(t-\tau_\ell)\ba(\theta_{\ell,p})+\bw_\mathrm{dl}(t)
\end{align}
where $\by(t,f_\mathrm{dl})=[y_1(t,f_\mathrm{dl}),\dots,y_J(t,f_\mathrm{dl})]^\mathrm{T}$, $\bS_\mathrm{dl}(t)=[\bs_{\mathrm{dl},1}(t),$ $\dots,\bs_{\mathrm{dl},J}(t)]$, and $\bw_\mathrm{dl}(t)=[w_{\mathrm{dl},1}(t),\dots,w_{\mathrm{dl},J}(t)]^\mathrm{T}$. Using the transformation in \eqref{channel model2}, the received signal in the frequency domain is represented as
\begin{align}\label{kOFDM}
\underline{\by}_k(f_\mathrm{dl})=\cF_k\{\by(t,f_\mathrm{dl})\}=\underline{\bS}_{\mathrm{dl},k}^\mathrm{H}\underline{\bh}_k(f_\mathrm{dl})+\underline{\bw}_{\mathrm{dl},k}.
\end{align}
Using the least square (LS) estimator based on the known $\underline{\bS}_{\mathrm{dl},k}$, the estimated DL CSI is given as
\begin{align}\label{estiDL}
\underline{\hat{\bh}}_k(f_\mathrm{dl})=(\underline{\bS}_{\mathrm{dl},k}^\mathrm{H})^{-1}\underline{\by}_k(f_\mathrm{dl})=\underline{\bh}_k(f_\mathrm{dl})+(\underline{\bS}_{\mathrm{dl},k}^\mathrm{H})^{-1}\underline{\bw}_{\mathrm{dl},k}.
\end{align}
Repeating the above process for all subcarriers, the DL OFDM channel $\hat{\underline{\bH}}(f_\mathrm{dl})$ can be obtained.

The DL training needs to be performed for every coherence time interval, and the training overhead normally increases as the number of antennas at the BS increases. Considering the training overhead, the effective spectral efficiency is given as
\begin{align}\label{eff_rate}
\mathfrak{R}_\text{eff}=\left(1-\frac{\text{training overhead}}{\text{coherence block length}}\right)\times \mathfrak{R}.
\end{align}
In \eqref{eff_rate}, $\mathfrak{R}$ is the spectral efficiency, which is defined as \cite{hassibi2003much}
\begin{align}\label{comm_rate}
\mathfrak{R}=\mathbb{E}\left[\frac{1}{K}\sum_{k=1}^K\log\left(1+\rho\left|\frac{\hat{\underline{\bh}}_k^\mathrm{H}(f_\mathrm{dl})}{\lVert\hat{\underline{\bh}}_k(f_\mathrm{dl})\rVert_2}\underline{\bh}_k(f_\mathrm{dl})\right|^2\right)\right],
\end{align}
where $\rho$ is the signal-to-noise ratio (SNR). We assume the perfect UL feedback in \eqref{comm_rate} to check the upper bound of the DL training-based transmission. The coherence time interval decreases as the speed of the MS increases, so the conventional DL-training-based approach is not appropriate for high mobility scenarios in FDD.

\subsection{NN-Based DL Extrapolation}\label{DLnn}

The purpose of using the NN is to estimate the DL CSI directly from the UL CSI measured at the BS without having any explicit downlink training. More precisely, the previous works in \cite{Alrabeiah:2019, Yang:2019} used the UL OFDM channel $\underline{\bH}(f_\mathrm{ul})$ as an input and the DL OFDM channel $\underline{\bH}(f_\mathrm{dl})$ as an output to train the NN, i.e.,
\begin{align}\label{full_ch}
\underline{\bH}(f_\mathrm{dl})=\cQ_\mathrm{CH}(\underline{\bH}(f_\mathrm{ul})),
\end{align}
where $\cQ_\mathrm{CH}(\cdot)$ denotes the function of NN-based DL extrapolation. We will denote this approach as ``CH-learning'' in the remaining of paper. Since the numbers of antennas at the BS and subcarriers in OFDM could be quite large, the CH-learning may suffer from heavy computational complexity.

\begin{figure}
	\centering
	\includegraphics[width=1.05\columnwidth]{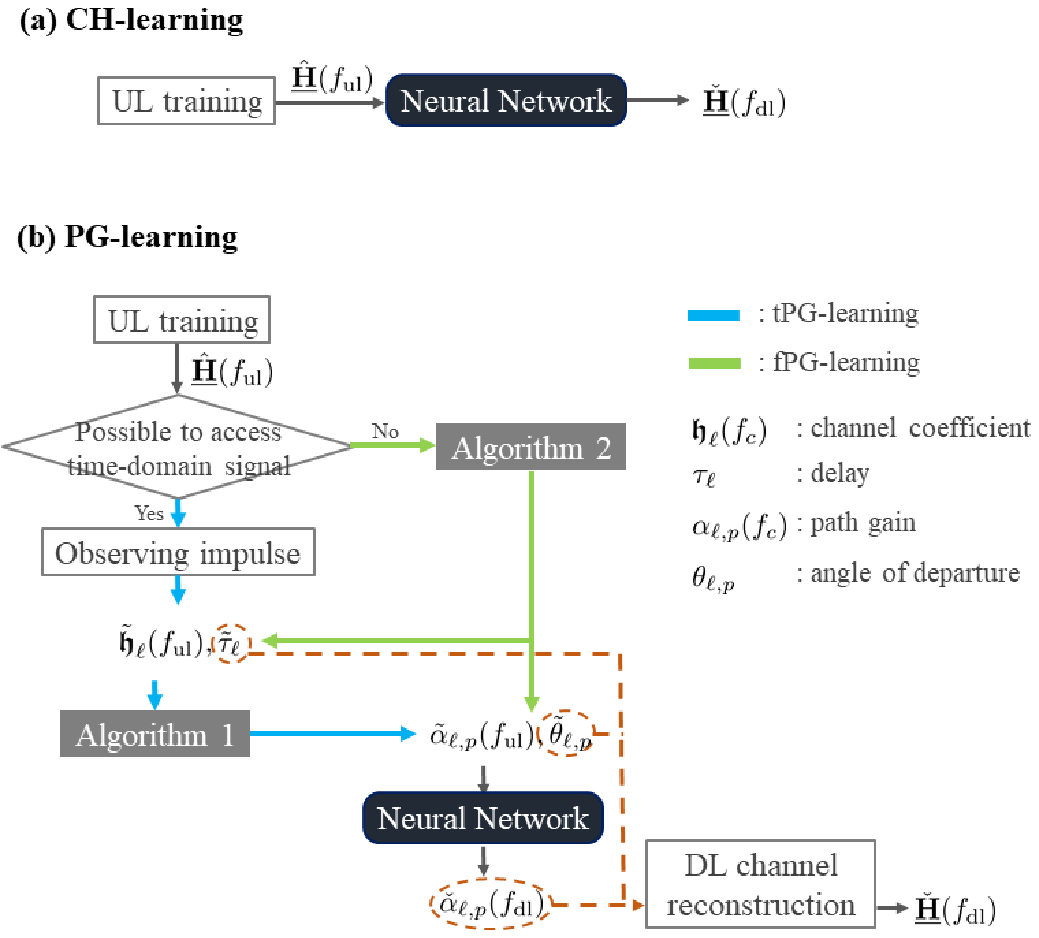}
	\caption{Flow charts of (a) CH-learning and (b) PG-learning.}
	\label{InSiteChannel}
\end{figure}

\section{Proposed NN-Based DL Extrapolation}\label{Param extrac}

In \eqref{channel model1}, the channel of $p$-th subpath from $\ell$-th cluster consists of three parameters, i.e., the path gain $\alpha_{\ell,p}(f_c)$, the delay $\tau_\ell$, and the AoD $\theta_{\ell,p}$. Taking the geometrical reciprocity into account, $\tau_\ell$ and $\theta_{\ell,p}$ estimated from the UL CSI would be identical to those of the DL channel. Therefore, instead of using the full OFDM channels, it would be possible to exploit only the path gains to train the NN for DL extrapolation. In Section \ref{PG learning}, we first explain a general procedure of the proposed DL extrapolation, which includes the process of DL channel reconstruction and the concept of proposed NN. Since the BS exploits the UL CSI for the DL extrapolation, we elaborate the UL training in Section \ref{ULtrain}. Then, we describe the algorithms to extract necessary information from the UL CSI depending on the availability of time domain information. If the BS can directly access the time domain CSI, we propose ``tPG-learning,'' which is detailed in Section \ref{t-PG}. If the BS only can access the frequency domain CSI, we introduce ``fPG-learning,'' which is covered in Section \ref{f-PG}.

\subsection{General Procedure of Proposed DL Extrapolation}\label{PG learning}

As explained in Remark 1, the DL extrapolation technique estimates the DL CSI at time $t_0+d$ using the UL CSI at time $t_0$. To simplify the notation, we intentionally neglect the delay $d$ in the following discussions.

In our proposed NN-based DL extrapolation, the DL channel can be reconstructed as 
\begin{align}\label{extrapolated channel}
\check{\bh}_\ell(t,f_\mathrm{dl})=\sum_{p=1}^R\check{\alpha}_{\ell,p}(f_\mathrm{dl})\delta(t-\tilde{\tau}_\ell)\ba(\tilde{\theta}_{\ell,p}),
\end{align}
where the delay $\tilde{\tau}_\ell$ and AoD $\tilde{\theta}_{\ell,p}$ are the extracted parameters from the UL CSI, which will be explained in Sections \ref{t-PG} and \ref{f-PG}. The DL path gain $\check{\alpha}_{\ell,p}(f_\mathrm{dl})$ is the parameter predicted from the extracted UL path gain $\tilde{\alpha}_{\ell,p}(f_\mathrm{ul})$ using the NN, i.e.,
\begin{align}\label{pglearning}
\{\check{\alpha}_{\ell,p}(f_\mathrm{dl})\}^{L,Q}=\cQ_\mathrm{PG}(\{\tilde{\alpha}_{\ell,p}(f_\mathrm{ul})\}^{L,Q}).
\end{align}
The function $\cQ_\mathrm{PG}(\cdot)$ in \eqref{pglearning} represents the network of the PG-learning. The detailed network structures of the PG-learning are explained in Section \ref{usedNNdetail}.

We use the tilde $\tilde{a}$ to denote extracted parameters from the UL CSI and the check $\check{a}$ to represent predicted parameters using the NN throughout the paper. In \eqref{extrapolated channel}, $R$ is the number of subpaths used for the channel reconstruction, $R\leq Q$, and $Q$ in \eqref{pglearning} is the number of subpaths used for the learning, $Q\leq P$. The effects of $R$ and $Q$ are explained in Sections \ref{usedNNdetail} and \ref{effect of P}. Using \eqref{channel model2}, the reconstructed DL channel in the time domain in \eqref{extrapolated channel} can be converted to an OFDM channel.

\subsection{UL Training}\label{ULtrain}
The BS obtains the UL CSI through the UL training. When the MS transmits a pilot symbol $s_\mathrm{ul}(t)$ to the BS, the received signal at the BS in the time domain is written as
\begin{align}
\by^\mathrm{H}(t,f_\mathrm{ul})=\sum_{\ell=1}^L\sum_{p=1}^P\alpha_{\ell,p}(f_\mathrm{ul})\ba^\mathrm{H}(\theta_{\ell,p})s_\mathrm{ul}(t-\tau_\ell)+\bw_\mathrm{ul}^\mathrm{H}(t).
\end{align}
The $N_\mathrm{BS}\times 1$ vector $\bw_\mathrm{ul}(t)$ is an additive white Gaussian noise with zero mean and variance $N_0\bI$. After the transformation, the received signal at the $k$-th subcarrier can be obtained as
\begin{align}\label{kRx}
\underline{\by}_k^\mathrm{H}(f_\mathrm{ul})=\cF_k\left\{\by^\mathrm{H}(t,f_\mathrm{ul})\right\} = \underline{\bh}_k^\mathrm{H}(f_\mathrm{ul})\underline{s}_{\mathrm{ul},k}+\underline{\bw}_{\mathrm{ul},k}^\mathrm{H},
\end{align} 
where the pilot symbol at the $k$-th subcarrier is $\underline{s}_{\mathrm{ul},k}=\cF_k\{s_\mathrm{ul}(t)\}$. The power of the transmitted symbol is constrained to $|\underline{s}_{\mathrm{ul},k}|^2\leq P_\mathrm{ul}$ for all $k$. The $N_\mathrm{BS}\times1$ noise vector $\underline{\bw}_{\mathrm{ul},k}=$ $\cF_k\{\bw_\mathrm{ul}(t)\}$ still maintains the same mean and variance as in the time domain case.
For the known symbol $s_\mathrm{ul}(t)$, the BS estimates the UL channel using the LS estimator
\begin{align}\label{UL_LS}
\hat{\underline{\bH}}^\mathrm{H}(f_\mathrm{ul}) = \underline{\bS}_\mathrm{ul}^{-1}\underline{\bY}^\mathrm{H}(f_\mathrm{ul}) = \underline{\bH}^\mathrm{H}(f_\mathrm{ul})+\underline{\bS}_\mathrm{ul}^{-1}\underline{\bW}_\mathrm{ul},
\end{align}
where  $\underline{\bY}(f_\mathrm{ul})=[\underline{\by}_1(f_\mathrm{ul})\cdots\underline{\by}_K(f_\mathrm{ul})]$, $\underline{\bS}_\mathrm{ul}=\diag[\underline{s}_{\mathrm{ul},1},$ $\dots,\underline{s}_{\mathrm{ul},K}]$, and $\underline{\bW}_\mathrm{ul}=[\underline{\bw}_{\mathrm{ul},1}\cdots\underline{\bw}_{\mathrm{ul},K}]$.

If the BS can access the time domain CSI, it is possible to directly obtain the noise corrupted $\ell$-th channel coefficient $\hat{\boldsymbol{\mathfrak{h}}}_\ell(f_\mathrm{ul})$ and delay $\hat{\tau}_\ell$ by observing impulses on the signal. If the BS only can access the OFDM channel $\hat{\underline{\bH}}(f_\mathrm{ul})$, however, the BS has to figure out $\boldsymbol{\mathfrak{h}}_\ell(f_\mathrm{ul})$ and $\tau_\ell$. We consider two cases: 1) tPG-learning: the case when the BS can directly access the time domain signals, and 2) fPG-learning: the case when the BS only can access the frequency domain signals, which are elaborated in the following subsections.

\begin{algorithm}[t]
	\caption{tPG-learning}
	\label{coeff extraction}
	\begin{algorithmic}
		\State \textbf{Initialization}: set $\bh_\ell^{(0)}=\hat{\boldsymbol{\mathfrak{h}}}_\ell(f_\mathrm{ul})$
		\State \textbf{for} $p = 1 : P$ \textbf{do}
		
		\begin{algorithmic}[]
			\State Maximizing inner product
			\begin{align}\tilde{\theta}_{\ell,p}=\operatorname*{arg\,max}_\theta\lVert\ba(\theta)^\mathrm{H}\bh_\ell^{(p-1)}\rVert^2
			\end{align}
			\State Extracting path gain
			\begin{align}
			\tilde{\alpha}_{\ell,p}(f_\mathrm{ul})=\frac{\ba(\tilde{\theta}_{\ell,p})^\mathrm{H}\bh_\ell^{(p-1)}}{\lVert\ba(\tilde{\theta}_{\ell,p})\rVert^2}
			\end{align}
			\State Nullspcae projection
			\begin{align}\bh_\ell^{(p)}=\bh_\ell^{(p-1)}-\tilde{\alpha}_{\ell,p}(f_\mathrm{ul})\ba(\tilde{\theta}_{\ell,p})
			\end{align}
		\end{algorithmic}
	\end{algorithmic}
	
	\begin{algorithmic}
		\State \textbf{end}
		\State \textbf{return}: $\{\tilde{\theta}_{\ell,p}\}^P$ and $\{\tilde{\alpha}_{\ell,p}(f_\mathrm{ul})\}^P$	
	\end{algorithmic}
	
\end{algorithm}

\subsection{\lowercase{t}PG-Learning}\label{t-PG}

In the tPG-learning, since the BS already has the estimated delay $\hat{\tau}_\ell$ and channel coefficient $\hat{\boldsymbol{\mathfrak{h}}}_\ell(f_\mathrm{ul})$, the BS only has to extract the AoDs and path gains.

Algorithm \ref{coeff extraction} shows how to extract the AoDs and path gains for the tPG-learning. In the beginning of the algorithm, the estimated channel coefficient $\hat{\boldsymbol{\mathfrak{h}}}_\ell(f_\mathrm{ul})$ is set to be $\bh_\ell^{(0)}=\hat{\boldsymbol{\mathfrak{h}}}_\ell(f_\mathrm{ul})$. The AoD of the first subpath can be extracted as 
\begin{align}\tilde{\theta}_{\ell,1}=\operatorname*{arg\,max}_\theta\lVert\ba(\theta)^\mathrm{H}\bh_\ell^{(0)}\rVert^2.
\end{align}
After finding the AoD of the first subpath $\tilde{\theta}_{\ell,1}$ , the path gain $\tilde{\alpha}_{\ell,1}(f_\mathrm{ul})$ corresponding to $\tilde{\theta}_{\ell,1}$ can be obtained as 
\begin{align}
\tilde{\alpha}_{\ell,1}(f_\mathrm{ul})=\frac{\ba(\tilde{\theta}_{\ell,1})^\mathrm{H}\bh_\ell^{(0)}}{\lVert\ba(\tilde{\theta}_{\ell,1})\rVert^2}.
\end{align}

To find the parameters of other subpaths accurately, following a similar concept in \cite{JSChoi:2019}, the effect of the dominant subpath could be cancelled by projecting $\bh_\ell^{(0)}$ to the nullspace of $\ba(\tilde{\theta}_{\ell,1})$.

\begin{align}\label{nullspace projection}
\underline{\bh}_\ell^{(1)}=\boldsymbol{\cP}_{\ba(\tilde{\theta}_{\ell,1})}^\perp\underline{\bh}_\ell^{(0)},
\end{align}
where $\boldsymbol{\cP}_{\ba(\tilde{\theta}_{\ell,1})}^\perp$ is the projection matrix to the nullspace of $\ba(\tilde{\theta}_{\ell,1})$, defined as
\begin{align}\label{nullspace projection}
\boldsymbol{\cP}_{\ba(\tilde{\theta}_{\ell,1})}^\perp=\bI_N-\frac{\ba(\tilde{\theta}_{\ell,1})\ba(\tilde{\theta}_{\ell,1})^\mathrm{H}}{\lVert\ba(\tilde{\theta}_{\ell,1})\rVert^2}.
\end{align}
As replacing $\bh_\ell^{(0)}$ with $\bh_\ell^{(1)}$ and repeating the same process,the parameters of second subpath; $\tilde{\theta}_{\ell,2}$ and $\tilde{\alpha}_{\ell,2}(f_\mathrm{ul})$, can be found. All the parameters constituting $\hat{\boldsymbol{\mathfrak{h}}}_\ell(f_\mathrm{ul})$ can be inferred through $P$ iterations.

\begin{algorithm}[t]
	\caption{fPG-learning}
	\label{delay extraction}
	\begin{algorithmic}
		\State \textbf{Initialization}: set $\underline{\bH}^{(0)}=\hat{\underline{\bH}}(f_\mathrm{ul})$
		\State \textbf{for} $\ell = 1 : L$ \textbf{do}
		
		\begin{algorithmic}[]
			\State 1. Maximizing inner product
			\begin{align}\label{eq:maxima}\tilde{\tau}_\ell=\operatorname*{arg\,max}_\tau\lVert\underline{\bH}^{(\ell-1)}\bp^\ast(\tau)\rVert_2^2
			\end{align}
			\State 2. Extracting channel coefficient
			\begin{align}\label{eq:projectedCH}
			\bar{\boldsymbol{\mathfrak{h}}}_\ell(f_\mathrm{ul})=\frac{\underline{\bH}^{(\ell-1)}\bp^\ast(\tilde{\tau}_\ell)}{\lVert\bp(\tilde{\tau}_\ell)\rVert^2}
			\end{align}
			\State 3. AoD and path gain extraction
			\\ \quad Extracting $\{\tilde{\theta}_{\ell,p}\}^P$ and $\{\tilde{\alpha}_{\ell,p}(f_\mathrm{ul})\}^P$ from $\bar{\boldsymbol{\mathfrak{h}}}_\ell(f_\mathrm{ul})$ \\ \quad using \textbf{Algorithm 1}\\
			\State 4. Approximating channel coefficient
			\begin{align}
			\bar{\boldsymbol{\mathfrak{h}}}_\ell(f_\mathrm{ul})\approx\tilde{\boldsymbol{\mathfrak{h}}}_\ell(f_\mathrm{ul})=\sum_{p=1}^P\tilde{\alpha}_{\ell,p}(f_\mathrm{ul})\ba(\tilde{\theta}_{\ell,p})
			\end{align}
			\State 5. Nullspcae projection
			\begin{align}\label{eq:Nullproj}\underline{\bH}^{(\ell)}=\underline{\bH}^{(\ell-1)}-\tilde{\boldsymbol{\mathfrak{h}}}_\ell(f_\mathrm{ul})\bp^\mathrm{T}(\tilde{\tau}_\ell)
			\end{align}
		\end{algorithmic}
	\end{algorithmic}
	
	\begin{algorithmic}
		\State \textbf{end}
		\State \textbf{return}: $\{\tilde{\theta}_{\ell,p}\}^{L,P}$, $\{\tilde{\alpha}_{\ell,p}(f_\mathrm{ul})\}^{L,P}$, $\{\tilde{\tau}_\ell\}^L$, and $\{\tilde{\boldsymbol{\mathfrak{h}}}_\ell(f_\mathrm{ul})\}^L$	
	\end{algorithmic}
	
\end{algorithm}

\subsection{\lowercase{f}PG-Learning}\label{f-PG}
In the fPG-learning, since the time domain CSI is not available, the BS should first estimate the delays and channel coefficients before extracting the AoDs and path gains, as detailed in Algorithm \ref{delay extraction}. Similar to Algorithm \ref{coeff extraction}, the delay $\tau_\ell$ can be obtained through the inner product maximization as in \eqref{eq:maxima}. After extracting $\tau_\ell$, projecting the estimated UL OFDM channel $\hat{\underline{\bH}}(f_\mathrm{ul})$ to $\bp(\tilde{\tau}_\ell)$ gives the channel coefficient $\bar{\boldsymbol{\mathfrak{h}}}_\ell(f_\mathrm{ul})$ in \eqref{eq:projectedCH}. Since all channel coefficients $\boldsymbol{\mathfrak{h}}_\ell(f_\mathrm{ul})$ are composed of the AoDs $\theta_{\ell,p}$ and path gains $\alpha_{\ell,p}(f_\mathrm{ul})$, they should be in the form of $\boldsymbol{\mathfrak{h}}_\ell(f_c)=\sum_{p=1}^P\alpha_{\ell,p}(f_c)\ba(\theta_{\ell,p})$. The channel coefficient $\bar{\boldsymbol{\mathfrak{h}}}_1(f_c)$ obtained in \eqref{eq:projectedCH}, however, may not follow such from since $\bar{\boldsymbol{\mathfrak{h}}}_1(f_c)$ is obtained from the noisy UL channel $\hat{\underline{\bH}}(f_\mathrm{ul})$. Therefore, with the extracted AoDs $\{\tilde{\theta}_{\ell,p}\}^P$ and path gains $\{\tilde{\alpha}_{\ell,p}(f_\mathrm{ul})\}^P$ in Step 3 of Algorithm \ref{delay extraction}, we approximate $\bar{\boldsymbol{\mathfrak{h}}}_1(f_c)$ as
\begin{align}\label{approxCHcoeff}
\bar{\boldsymbol{\mathfrak{h}}}_1(f_\mathrm{ul})\approx\tilde{\boldsymbol{\mathfrak{h}}}_1(f_\mathrm{ul})=\sum_{p=1}^P\tilde{\alpha}_{1,p}(f_\mathrm{ul})\ba(\tilde{\theta}_{1,p}).
\end{align}
With the approximated $\tilde{\boldsymbol{\mathfrak{h}}}_1(f_\mathrm{ul})$, the nullspace projection, which is necessary for finding the other channel parameters, is conducted in \eqref{eq:Nullproj}. All the channel parameters from $L$ clusters can be obtained by repeating the process $L$ times. 

\begin{figure*}
	\centering
	\includegraphics[width=1.7\columnwidth]{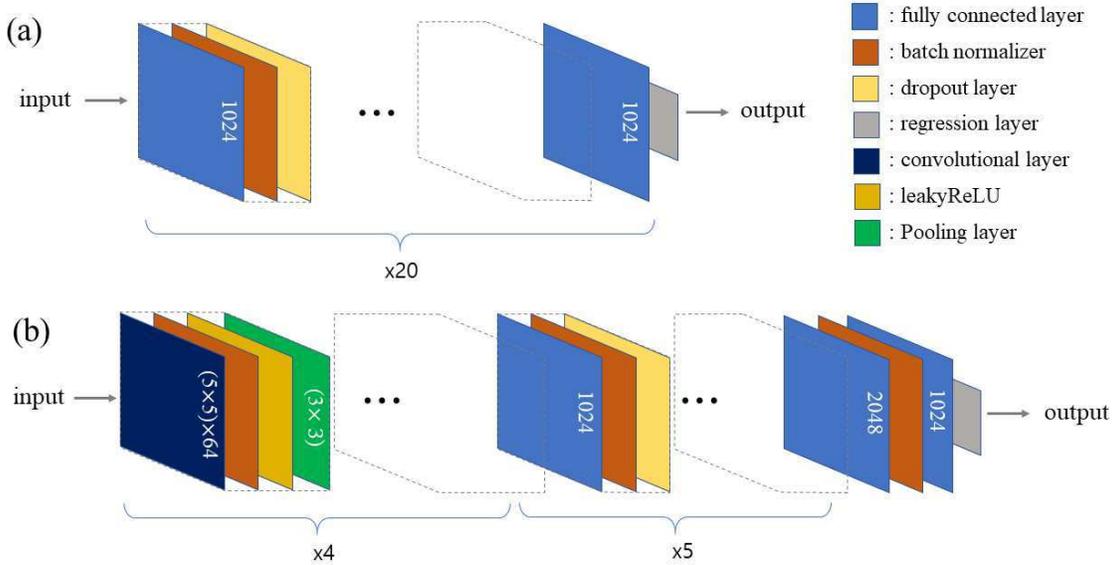}
	\caption{NN structures used for numerical studies. (a) MLP for the CH-learning and (b) CNN for the PG-learning.}
	\label{NNstructure}
\end{figure*}

\section{Deep Learning Settings}\label{NN Settings}

When using the NN, there are many things to consider to optimize its performance, e.g., pre-processing data sets, designing an NN structure, and setting hyper parameters. We have performed extensive simulations to find the best settings for the CH-learning and the PG-learning.

In this section, we first discuss the structure of input and output layers of the NN. Then, we explain the reason why we choose multilayer perceptron (MLP) for the CH-learning and convolutional neural network (CNN) for the proposed PG-learning. Finally, we elaborate the NN structures for the CH and PG-learning.

\subsection{Input and Output Layers for NN}\label{inoutNN}

The OFDM channels and extracted channel path gains, which are the inputs and outputs of the NN for the CH-learning and proposed PG-learning, respectively, are all complex-valued data. While there are some recent work on NNs that can deal with complex-valued inputs and outputs \cite{Xiao:2019,Yang:2019}, we exploit well-established NNs using real-valued data.

To convert complex-valued data into real-valued ones, the most commonly used method is to divide the complex-valued data into their real and imaginary parts. The real and imaginary parts may form one input layer \cite{Arn:2019,Dong:2019,Huang:2019,Wang:2019,Alrabeiah:2019,Navabi:2018}, or each of parts can construct two independent input layers \cite{Wen:2018}, where each input layer is used for a separate network. Not only the real and imaginary parts of data but also the magnitude of data can be added to the input layer as in \cite{Elbir:2019}, since more information, even redundant, in the input layer restricts the network from over-fitting. Depending on the experimental scenarios, we select proper structures of the input and output layers that give the best performance.

\subsection{Proper NN Selection}\label{MLP}

Many previous works on wireless communication systems tried to solve nonlinear problems with MLP and CNN. MLP is the most common form of the NN. Some works on the DL extrapolation used MLP \cite{Arn:2019,Yang:2019,Alrabeiah:2019,Navabi:2018}. CNN is another NN that is superior for pattern recognition \cite{geron2019hands}. CNN has also been used for the estimation of wireless channels or channel parameters \cite{Viera:2017,Arn:2019,Dong:2019}.

The optimal NN structure greatly depends on the data, so the CH-learning and the PG-learning may require different structures. The CH-learning based on CNN and MLP is studied in \cite{Arn:2019} and \cite{Alrabeiah:2019}, respectively. After extensive simulations, we have verified that MLP is better than CNN for the CH-learning while CNN outperforms MLP for the PG-learning.

The CH-learning uses the full dimensional channels directly as the input and output of NN, where the channel parameters are superposed, and a pattern within the input and output channel is vague, making MLP, which works well with highly uncertain data \cite{Driss:2017}, suitable. CNN is more proper to the pattern recognition, which works well with more structural data. The PG-learning operates with the path gains extracted from the full dimensional channels. The path gains are more structured than the full dimensional channels, which makes CNN more preferable to MLP for the PG-learning. 

\subsection{Details of NN Structures}\label{usedNNdetail}

The structure of NN we used for the CH-learning is shown in Fig. \ref{NNstructure}-(a). MLP in Fig. \ref{NNstructure}-(a) consists of 20 units assembled with a fully connected layer, a batch normalizer, and a dropout layer. Performance of MLP is much improved when 20 units or more are used. In the CH-learning, the activation function is omitted because it worsens the result. In Fig. \ref{NNstructure}-(a), the number written on the fully connected layer represents the number of nodes on the layer. The number of weights to be updated in Fig. \ref{NNstructure}-(a) is $((N_\mathrm{BS}\times K)\times1024)\times2+(1024/2)^{2}\times19+1024^2=6029312+2048\times(N_\mathrm{BS}\times K)$. The size of the input and output is $N_\mathrm{BS}\times K$, which can greatly increase the size of the network.

The CNN structure used for the PG-learning is shown in Fig. \ref{NNstructure}-(b). The front part of CNN consists of four convolutional units, which include a convolutional layer, a batch normalizer, an activation function, and a pooling layer. By using the activation function and the pooling layer, the effect of feature extraction is maximized. Among many possible activation function including the ReLU, clipped ReLU, leakyReLU, ELU, and hyper tangent function, it turned out that the leakyReLU gives the best performance for our data. At the rear of network, the same unit as those used in MLP is repeated five times, which is used to map features to the outputs.

In Fig. \ref{NNstructure}-(b), the number $(5\times 5)\times 64$ on the convolutional layer means that the number of convolutional kernels is 64, and its size is $5\times 5$, and $(3\times 3)$ on the pooling layer also means the kernel size. The number of weights to be updated for the PG-learning in Fig. \ref{NNstructure}-(b) is $((5\times 5)\times 64)\times4+((Q\times L)\times1024)\times2+(1024/2)^2\times4+(1024/2\times2048)+2048\times 1024=422704+2048\times(Q\times L)$ where $Q\times L$ is the size of the input and output of the NN. Note that $P$ subpath components are extracted by Algorithm \ref{coeff extraction} but not all $P$ path gains are used as the input and output for the PG-learning. By choosing $Q$ out of $P$ path gains, it is possible to balance between the learning complexity and NN performance. The effect of $Q$ is verified in Section \ref{effect of P}. Since $Q$ and $L$ are not related to the dimension of channels, the size of network in the PG-learning can be much smaller than the CH-learning, especially when the BS has a large number of antennas.

\begin{table}
	\caption{The adopted QuaDRiGa data-set parameters.}
	\label{table}
	\setlength{\tabcolsep}{3pt}
	\begin{center}
		\begin{tabular}{|p{115pt}|p{50pt}|}
			\hline
			Small Scale Fading Parameters & Value \\ \hline
			UL carrier frequency & 2.6 GHz \\
			DL carrier frequency & 2.9 GHz \\
			Total bandwidth & 100 MHz \\
			Coherence bandwidth & 180 kHz \\
			Tx. and Rx. antenna & `3gpp-3d' \\
			Number of clusters (paths) & 7 \\
			Number of subcarriers & 32 \\
			Transmit power (MS) & 30 dBm \\
			Noise variance ($N_0$) & -174 dBm \\
			Speed of MS & 10 km/h \\
			Decorrelation distance & 5 m \\
			\hline
		\end{tabular}
	\end{center}
	\label{QuaDRiGa parameters}
\end{table}

\section{Numerical Results}\label{Numerical result}


We use QuaDRiGa to generate realistic DL and UL channels to compare the proposed tPG and fPG-learning to the CH-learning and the conventional DL training. Parameters used in QuaDRiGa are listed in Table \ref{QuaDRiGa parameters}. Specific parameter settings are stated in each result if the parameters are different from Table \ref{QuaDRiGa parameters}.

We consider the environment with seven clusters as a typical case of an urban communication scenario. To mimic the blockage caused by variable obstacles and moving vehicles, we randomly change one to three clusters for each MS sample set. The snapshots of each MS sample set are obtained every 40 msec, which is a period of the UL sounding reference signals. We consider the processing delay $d=5$ msec caused in the process of the NN-based DL extrapolation, which is explained in Remark 1. For the training set, we generate 200 MS sample sets in total. The accuracy of DL extrapolation is measured by the correlation factor, $\rho_{\text{CORR}}$, which is defined as
\begin{align}\label{corr_fac}
\rho_\text{CORR} = \mathbb{E}\left[\frac{1}{K}\sum_{k=1}^K\frac{\lvert\underline{\bh}_k^\mathrm{H}(f_\mathrm{dl})\check{\underline{\bh}}_k(f_\mathrm{dl})\rvert}
{\lVert\underline{\bh}_k(f_\mathrm{dl})\rVert_2\lVert\check{\underline{\bh}}_k(f_\mathrm{dl})\rVert_2}\right].
\end{align}

In this section, we first check the effect of the number of subpaths on the tPG and fPG-learning in Section \ref{effect of P}. Then, we compare the tPG, fPG, and CH-learning in the following subsections. In Section \ref{comparison with Learning}, we verify the effect of the total bandwidth, UL transmit power, and speed of MSs. Finally, we compare the PG-learning to the CH-learning and the conventional DL training in Section \ref{comparison with training}. 

\begin{figure}
	\centering
	\includegraphics[width=0.95\columnwidth]{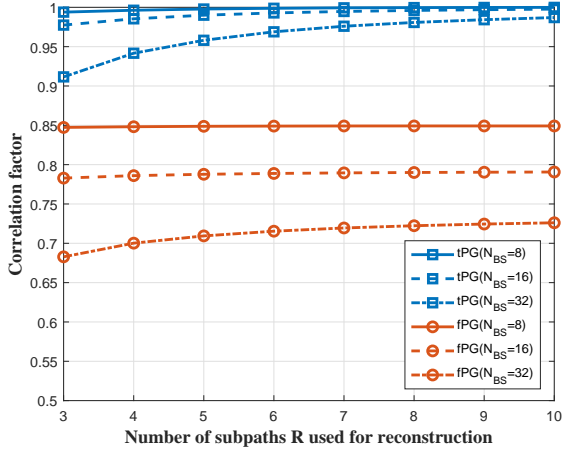}
	\caption{The correlation factors of the reconstructed channel with perfectly inferred path gains versus the number of subpaths $R$ used for the reconstruction with different values of $N_\mathrm{BS}$.}
	\label{recon_perfect}
\end{figure}

\subsection{Effect of Number of Subpaths on PG-Learning}\label{effect of P}

\begin{figure}
	\centering
	\includegraphics[width=1.05\columnwidth]{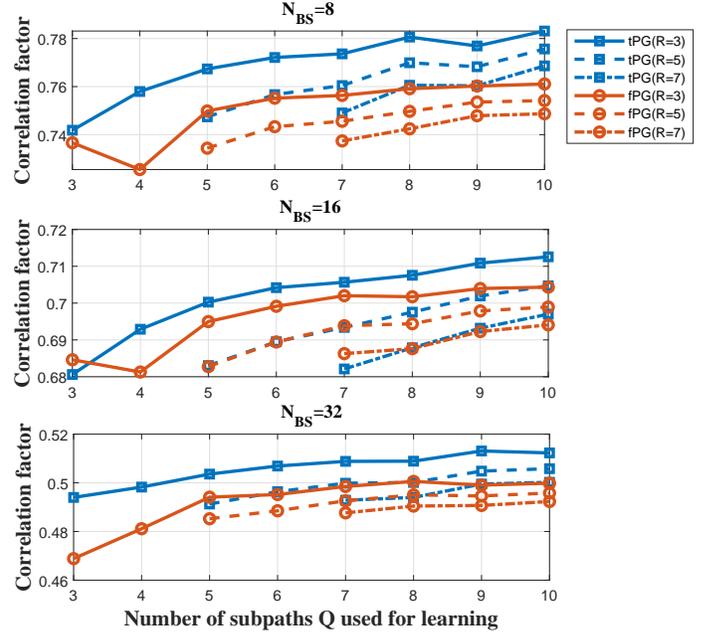}
	\caption{The correlation factors of the tPG and fPG-learning versus the number of subpaths $Q$ used for the learning with different values of $N_\mathrm{BS}$ and $R$.}
	\label{recon}
\end{figure}

It it necessary to examine the effect of $R$, the number of subpaths for the DL extrapolation explained in Section \ref{PG learning}, and $Q$, the number of subpaths for the PG-learning explained in Section \ref{usedNNdetail}. Fig. \ref{recon_perfect} shows the effect of $R$ with the perfectly inferred DL path gains. If the PG-learning could infer the exact DL path gains, the accuracy of the DL channel reconstruction increases with $R$. While using large $R$ seems to be helpful for the precise channel reconstruction as shown in Fig. \ref{recon_perfect}, with the predicted path gains (with errors) from the PG-learning, Fig. \ref{recon} shows the accuracy of reconstruction decreases as $R$ increases. This is because the errors in the predicted subpath components add up, making the reconstructed channel less accurate. This result points out that there is still room for improvement in the proposed NN-based DL extrapolation.

Fig. \ref{recon} also indicates the effect of $Q$ in the learning. The overall simulation results show that having large $Q$ improves the performance of the PG-learning, since the increased size of input and output puts more information to the network for the PG-learning with additional NN training overhead.

\begin{figure}
	\centering
	\includegraphics[width=1\columnwidth]{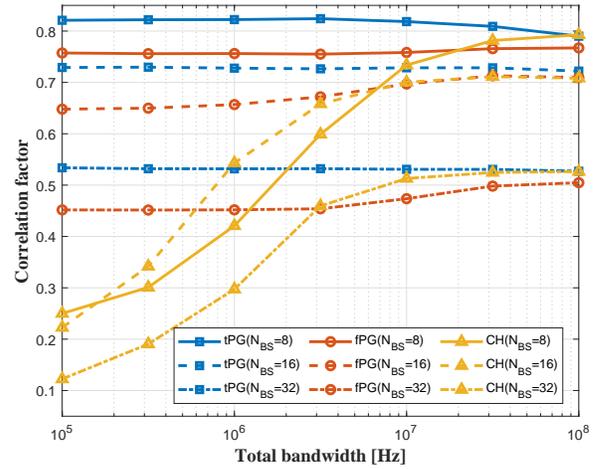}
	\caption{The correlation factors of the tPG, fPG, and CH-learning versus the total bandwidth with different values of $N_\mathrm{BS}$.}
	\label{sampF}
\end{figure}

\begin{figure}
	\centering
	\includegraphics[width=1\columnwidth]{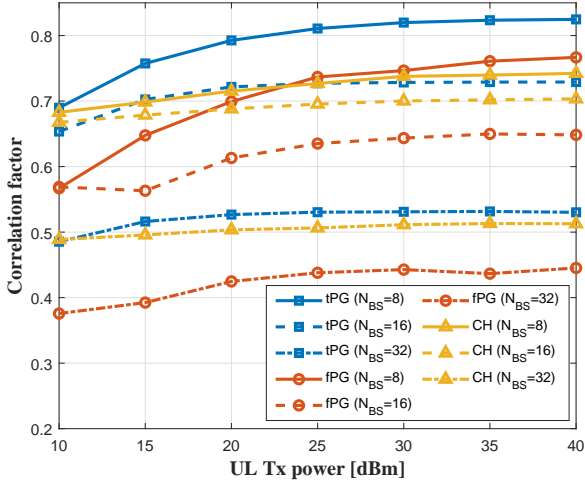}
	\caption{The correlation factors of the CH and PG-learning versus the UL transmit power with different values of $N_\mathrm{BS}$.}
	\label{SNR}
\end{figure}

\subsection{CH-Learning and PG-Learning}\label{comparison with Learning}

Tendency of the correlation factor according to the total bandwidth is shown in Fig. \ref{sampF}. As shown in Fig. \ref{sampF}, the CH-learning is seriously affected by the bandwidth. This is because the CH-learning is based on the OFDM channels, which vary with the total bandwidth significantly. The total bandwidth determines the phase difference between subcarriers as shown in \eqref{channel model2}. This phase difference is negligible when the bandwidth is small because the exponents in $\bp(\tau_\ell)$, which is defined after \eqref{OFDMchannel}, go to zero as $f_s\tau_\ell\approx 0$ with a small bandwidth, i.e., $\bp(\tau_\ell)\approx[1,\dots,1]^\mathrm{T}$ for all $\ell$. As a consequence, all channel coefficients $\boldsymbol{\mathfrak{h}}_\ell(f_c)$ overlap on the same base vector $\bp(\tau_\ell)$, which makes the CH-learning difficult to predict the DL OFDM channels.

\begin{figure}
	\centering
	\includegraphics[width=1.05\columnwidth]{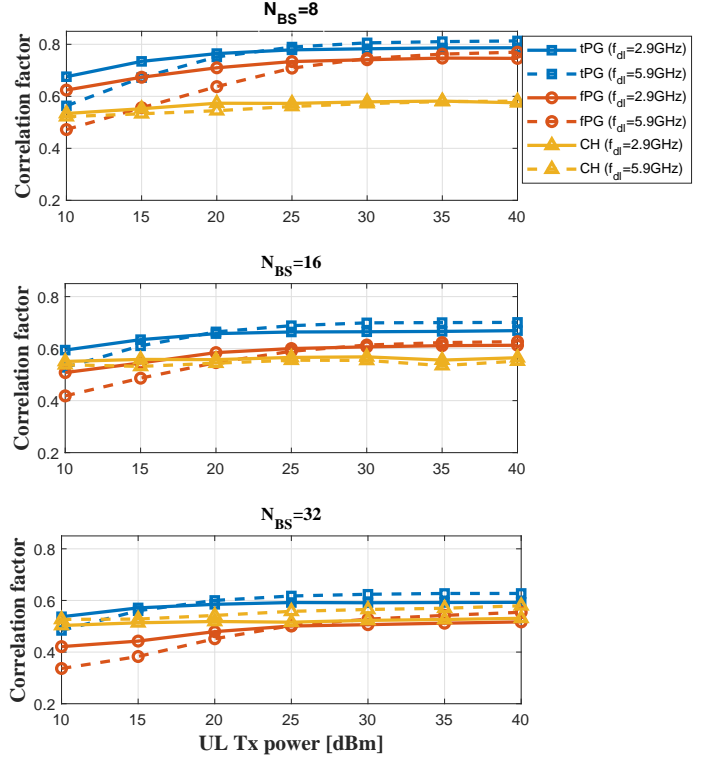}
	\caption{The correlation factors of the CH and PG-learning versus the UL transmit power with different values of $N_\mathrm{BS}$. Different values, (2.6, 2.9) and (5.9, 6.2) GHz for the (UL, DL) carrier frequencies, are considered.}
	\label{SNR_fc}
\end{figure}

The PG-learning does not use the OFDM channels in the learning phase but the correlation factor of the PG-learning is also measured on the transformed DL OFDM channels, which means that the PG-learning can be also affected by the total bandwidth. As shown in Fig. \ref{sampF}, however, the correlation factors of PG-learning are quite stable regardless of the bandwidth.

Fig. \ref{SNR} shows the effect of UL parameter estimation errors on the DL extrapolation. As shown in Fig. \ref{SNR}, the fPG-learning has more difficulty in the DL extrapolation than the tPG-learning at low UL transmit power. Since the fPG-learning extracts the path gains and AoDs from the extracted channel coefficients, it is more vulnerable to the UL parameter estimation errors than the tPG-learning. This result proves that the precise DL channel reconstruction highly depends on accurate UL channel estimates. The CH-learning, on the contrary, is not affected much from the UL transmit power since it just exploits the UL channel estimates without further processing, which makes more robust to the estimation error.

\begin{figure}
	\centering
	\includegraphics[width=1.05\columnwidth]{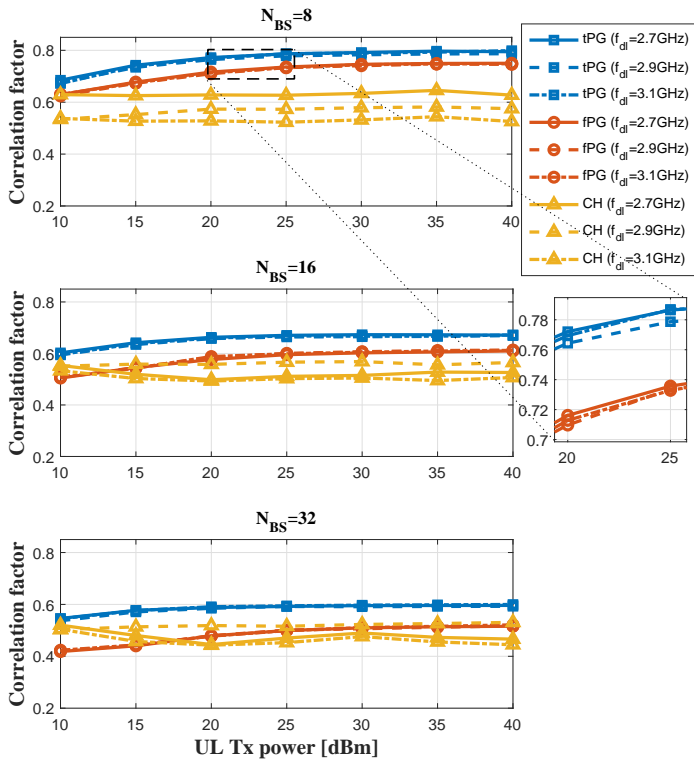}
	\caption{The correlation factors of the CH and PG-learning versus the UL transmit power with different values of $N_\mathrm{BS}$. To test different guard bands, three different values, (2.6, 2.7), (2.6, 2.9), and (2.6, 3.1) GHz for the (UL, DL) carrier frequencies are considered.}
	\label{SNR_bg}
\end{figure}

\begin{figure}
	\centering
	\includegraphics[width=1.05\columnwidth]{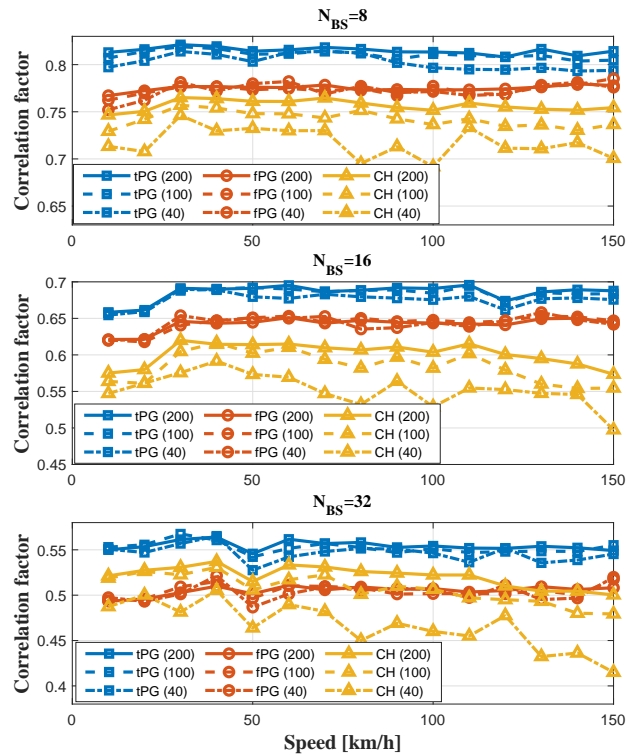}
	\caption{The correlation factors of the CH and PG-learning versus the MS speed with different values of $N_\mathrm{BS}$. We test three different values, 200, 100, and 40, for the number of MS sample sets. }
	\label{speed}
\end{figure}

\begin{figure*}
	\centering
	\includegraphics[width=1.60\columnwidth]{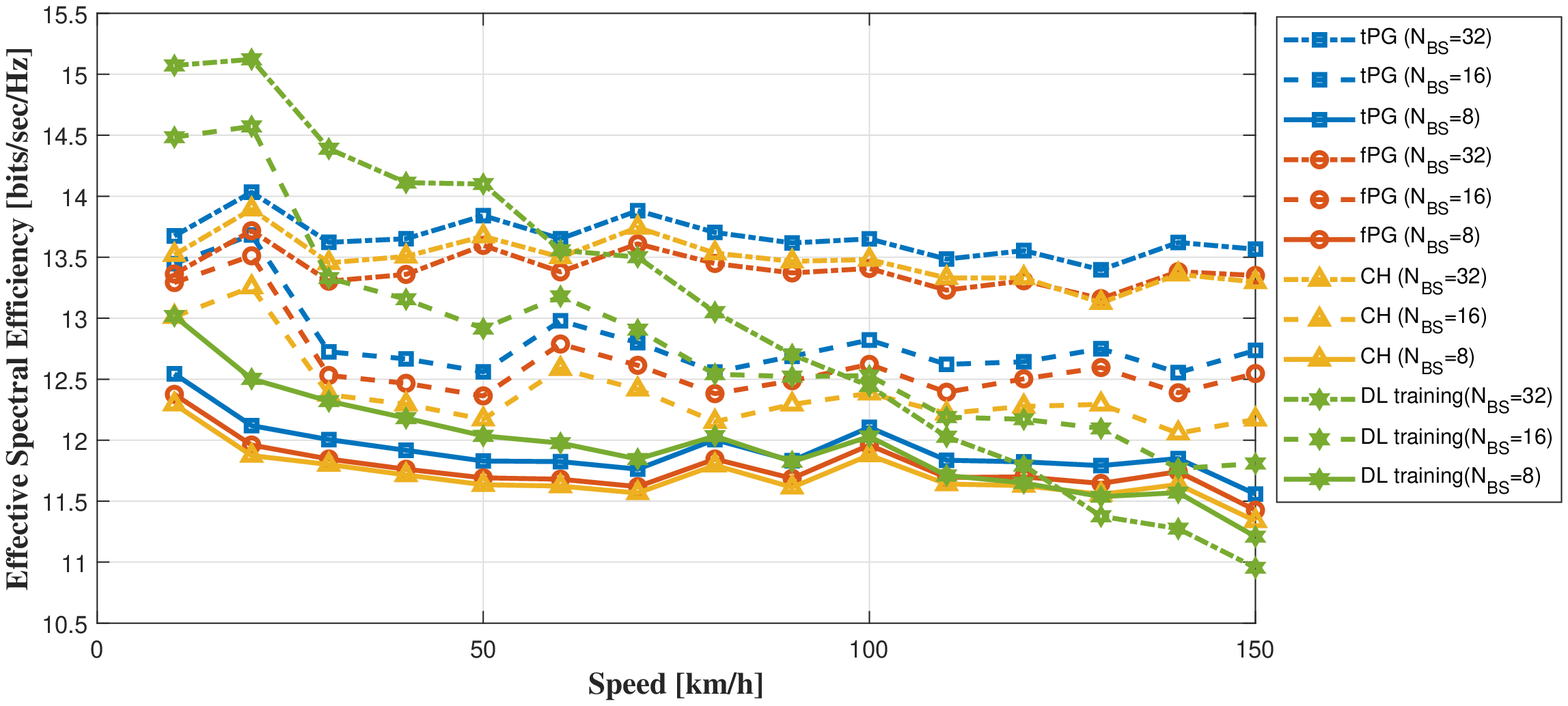}
	\caption{The effective rates of the CH-learning, the PG-learning, and the DL training versus the MS speed with different values of $N_\mathrm{BS}$.}
	\label{rate}
\end{figure*}

The carrier frequency and guard band can also affect the NN-based DL extrapolation. We evaluate with different carrier frequency in Fig. \ref{SNR_fc} and different guard bands in Fig. \ref{SNR_bg}. In \eqref{channel model2}, the path gain $\alpha(f_c)$ is the only channel parameter related to the carrier frequency. When the carrier frequency is high, Fig. \ref{SNR_fc} shows that the proposed PG learning requires to have more UL transmit power to work properly since the path gain $\alpha(f_c)$ attenuates more rapidly at high carrier frequency. However, Fig. \ref{SNR_bg} shows that the guard band seems to barely affect the DL extrapolation. Even for the 0.5 GHz guard band, which might be larger than most of practical FDD systems, the NN-based DL extrapolation methods work well as for the 0.1 GHz guard band. As the number of antennas at the BS increases, the overall correlation factors decrease but the tendency according to the guard band is maintained.

The correlation factors of the tPG, fPG, and CH-learning are compared with the MS speed in Fig. \ref{speed}. The overall results show that the accuracy decreases at high speed.\footnote{The simulation environments with different speeds are based on different cluster geometries while all three methods are tested with the same geometry for the same speed. Therefore, lower speeds do not always give higher correlation factors due to different cluster geometries. The result of each speed is averaged over ten different geometries.} When the MS speed is high, it may not be able for the MS to collect sufficient number of samples for the given length of the path since the snapshots of each MS sample set are generated every 40 msec. This problem can be resolved by increasing the number of MS sample sets. In Fig. \ref{speed}, we train each learning method with 40, 100, and 200 MS sample sets. With 40 MS sample sets, the CH-learning suffers from insufficient number of data for training. On the contrary, the PG-learning is robust to the lack of samples. In addition, the accuracy of PG-learning is quite stable even at high speed compared to the CH-learning, which proves that the PG-learning is easier to train.

Note that the channel of MS with high speed could be outdated easily during the processing delay $d$. Since we have already taken $d$ into account for the training, the processing delay does not affect the accuracy even at high speed.

\subsection{Comparison with DL Training}\label{comparison with training}

To compare the DL spectral efficiencies of the DL training and the NN-based DL extrapolation techniques, the coherence block length in \eqref{rate} is derived as
\begin{align}
\text{coherence block length}=f_\mathrm{coh}\times T_\mathrm{coh},
\end{align}
where $f_\mathrm{coh}$ is the coherence bandwidth, which is set to 180 kHz as in Table \ref{QuaDRiGa parameters}, and $T_\mathrm{coh}$ is the coherence time defined as
\begin{align}\label{eq:coherence time}
T_\mathrm{coh}=\frac{c}{4f_c v}.
\end{align}
In \eqref{eq:coherence time}, $c$ is the speed of light, $f_c$ is the DL carrier frequency, and $v$ is the speed of MS. The DL training overhead is $N_\mathrm{BS}$, and the transmit power at the BS is set to 30 dBm. For the DL training, we assume that the estimated DL CSI at the MS is perfectly fed back to the BS.

From Fig. \ref{rate}, it is clear that the DL training suffers from large training overhead at high speed due to the short coherence block length. The performance degradation of DL training is especially severe when the number of BS antennas, $N_\mathrm{BS}$, is large, making it unsuitable to massive MIMO. This is quite clear from \eqref{eff_rate}, i.e., with higher MS speed, the coherence block length decreases, and with more BS antennas, the training overhead increases. On the contrary, the NN-based DL extrapolation does not suffer from any DL training overhead. Moreover, as discussed in the previous subsection, the NN-based DL extrapolation does not suffer much for high MS speed by taking the processing delay into consideration for the training. The performance gap between the DL training and DL extrapolation would become even more prominent once we consider the limited feedback. Note that the proposed tPG-learning is always superior to the CH-learning while the fPG-learning and the CH-learning are comparable with each other.

\section{Conclusion}\label{conclusion}
In the communication environment with high mobility, the DL training method cannot work well, especially when the number of BS antennas becomes large, since the DL training causes large overhead and considerable delay. To overcome this problem, we proposed a novel DL extrapolation technique, dubbed as the PG-learning, using the UL CSI in this paper. Different from previous works on DL extrapolation relying on the full UL and DL channels, dubbed as the CH-learning, the proposed PG-learning exploits low-dimensional extracted channel paths gains as the input and output of NN. The numerical results showed that the PG-learning has higher DL extrapolation accuracy, regardless of the speed of MS, with significantly lower NN training overhead than the CH-learning.

We also verified, however, that there is still room to improve the PG-learning. More accurate UL path gain estimation and DL path gain prediction can lead to an advance on our scheme. Instead of relying on a discrete grid search for the AoD and delay as in this paper, the UL path gain estimation could be improved by gridless search using, e.g., atomic norm minimization \cite{chi2019harnessing}. To improve DL path gain prediction, channel tracking can be applied. The channel tracking can adapt various methods; Kalman filter (KF), Rauch-Tung-Striebel smoother (RTSS), and K-mean clustering \cite{Jianpeng2019,Muye2019,Jianwei2018,Yu2019}. The recursive neural network (RNN), e.g., LTSM \cite{Gui:2018}, can be also considered as a channel tracking method. With more precise UL and DL path gain information, the proposed NN-based DL extrapolation may be improved.

\end{document}